\documentclass[aip,graphicx, reprint]{revtex4-2}

\draft % marks overfull lines with a black rule on the right

\usepackage{graphicx}% Include figure files
\usepackage{dcolumn}% Align table columns on decimal point
\usepackage{bm}% bold math
\usepackage[utf8]{inputenc}
\usepackage[T1]{fontenc}
\usepackage{mathptmx}
\usepackage{etoolbox}
\usepackage{color}

\makeatletter
\def\@email#1#2{%
 \endgroup
 \patchcmd{\titleblock@produce}
  {\frontmatter@RRAPformat}
  {\frontmatter@RRAPformat{\produce@RRAP{*#1\href{mailto:#2}{#2}}}\frontmatter@RRAPformat}
  {}{}
}%
\makeatother
\begin{document}

%\preprint{}

\title{Top-down integration of a hBN quantum emitter in a monolithic photonic waveguide} %Title of paper



\author{Domitille G\'erard$^{1}$, Michael Rosticher$^{2}$, Kenji Watanabe$^3$, Takashi Taniguchi$^4$, Julien Barjon$^1$, St\'ephanie Buil$^1$, Jean-Pierre Hermier$^1$,  Aymeric Delteil$^{1,*}$}
%\thanks{}
\affiliation{$^1$ Universit\'e Paris-Saclay, UVSQ, CNRS,  GEMaC, 78000, Versailles, France. \\
$^2$ Laboratoire de Physique de l’École Normale Supérieure, ENS, Université PSL, CNRS, Sorbonne Université, Université de Paris, Paris, France. \\
$^3$ Research Center for Functional Materials, 
National Institute for Materials Science, 1-1 Namiki, Tsukuba 305-0044, Japan \\
$^4$ International Center for Materials Nanoarchitectonics, 
National Institute for Materials Science, 1-1 Namiki, Tsukuba 305-0044, Japan \\
$^*$ aymeric.delteil@uvsq.fr}


\date{\today}

\begin{abstract}
Integrated quantum photonics, with potential applications in quantum information processing, relies on the integration of quantum emitters into on-chip photonic circuits. Hexagonal boron nitride (hBN) is recognized as a material that is compatible with such implementations, owing to its relatively high refractive index and low losses in the visible range, together with advantageous fabrication techniques. Here, we combine hBN waveguide nanofabrication with the recently demonstrated local generation of quantum emitters using electron irradiation to realize a fully top-down elementary quantum photonic circuit in this material, operating at room temperature. This proof of principle constitutes a first step towards deterministic quantum photonic circuits in hBN.
\end{abstract}


\maketitle %\maketitle must follow title, authors, abstract and \pacs

Hexagonal boron nitride (hBN) has recently emerged as a very attractive platform for integrated quantum photonics~\cite{Wang19, Pelucci21}. This van der Waals (vdW) material offers a wide range of fabrication techniques that allow to associate it with other materials --including other vdW crystals-- in highly miniaturized complex devices. In particular, it presents favorable properties for photonics, with atomically flat surfaces and a very wide bandgap ($\sim 6$~eV), opening the possibility to use it as a light confining medium. In this spirit, fabrication of complex hBN photonic structures, such as waveguides~\cite{Li21,Khelifa22}, phase plates and microlenses~\cite{Lassaline21}, bullseye antennas~\cite{Froch21} and photonic crystal structures~\cite{Kim18,Froch18}, have been recently demonstrated.

Last but not least, hBN also hosts optically active point defects that act as excellent single-photon emitters (SPEs) in various wavelength ranges~\cite{Tran16,Bourrelier16,Martinez16}. Most of these color centers occur randomly in the flake, thereby hindering scalable integration in photonic devices. Nonetheless, these emitters have been at the core of highly promising implementations of both monolithic and hybrid photonic devices, including waveguides~\cite{Li21, Kim19, Elshaari21}, cavities~\cite{Kim18,Froch20,Parto22} and fibers\cite{Schell17,Vogl17,Haussler21}. Those realizations are relying on either \textit{a posteriori} integration of the quantum emitter, or on the random presence of an emitter in the structure, which limits both control and scalability of those devices.

The recent demonstration of local generation of blue-emitting color centers (B-centers) using a focused electron beam has offered an attractive workaround~\cite{Shevitski19, Fournier21, Gale22}. These emitters can be generated in a commercial scanning electron microscope (SEM) with a high control of their position and average number, and consistently exhibit a reproducible emission wavelength, a predominent in-plane polarization, a short lifetime and a high optical coherence~\cite{Fournier21, Gale22, Fournier22, Horder22, Fournier23}.

Here, we take advantage of this e-beam technique by including it in a completely top-down approach for the fabrication of an elementary quantum photonic device, where the emitter generation is included as an additional step in the fabrication process. We first fabricate short waveguides (10~$\mu$m) with semicircular grating couplers~\cite{Javadi15,Patil22} and subsequently embed quantum emitters in the waveguide by local irradiation. Photoluminescence (PL) characterization demonstrates the coupling of both the excitation laser and the SPE emission into the waveguide. Although the design we implemented is not intended to be optimal, it illustrates the potential of electron-beam generated SPEs for quantum photonics and integrated optical quantum information.

\begin{figure}
 \includegraphics[width=0.95\linewidth]{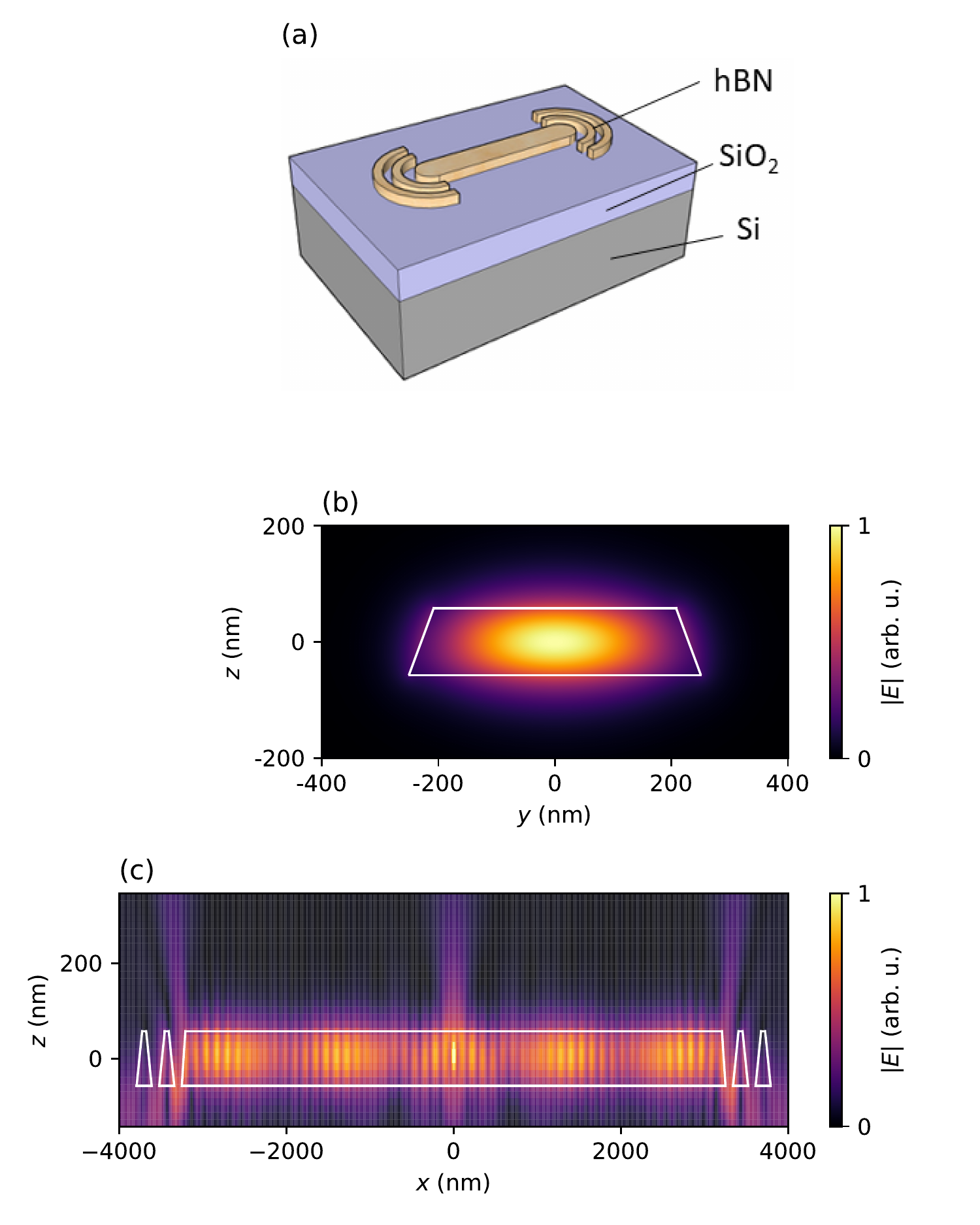}%
 \caption{\label{fig1} Design of the hBN waveguide embedding quantum emitters. (a) Scheme of the hBN waveguide on SiO$_2$/Si embedding a SPE. (b) TE$_{00}$ mode profile as calculated with FDTD. (c) Longitudinal cut of the dipole emission propagation in the structure as calculated with FDTD.}%
 \end{figure}

The geometry that we have opted for is a ridge waveguide, chosen for the simplicity of its realization. The light is confined by refractive index contrast between hBN ($n_o \sim 2.2$) and the environment. The SiO$_2$/Si substrate has a refractive index that is low enough to obtain low-losses propagating modes in flakes as thin as 60~nm. Fig.~\ref{fig1}(a) shows a sketch of the waveguide with semicircular grating couplers at its two output ports. Fig.~\ref{fig1}(b) shows the waveguide cross section and the corresponding FDTD simulation of the fundamental TE mode profile. Fig.~\ref{fig1}(c) shows the longitudinal profile of the same mode. For a point dipole emitting at 440~nm with an in-plane polarization orthogonal to the waveguide main axis and located at the mode antinode, we calculate that 23~\% of the light is coupled to the waveguide in each direction, of which 18~\% is extracted towards the top direction to be collected by a NA~=~0.8 lens. Additionally, 5~\% is directly coupled to the upper free space, allowing to characterize the sample without using the guided modes.
 
 \begin{figure}
 \includegraphics[width=0.85\linewidth]{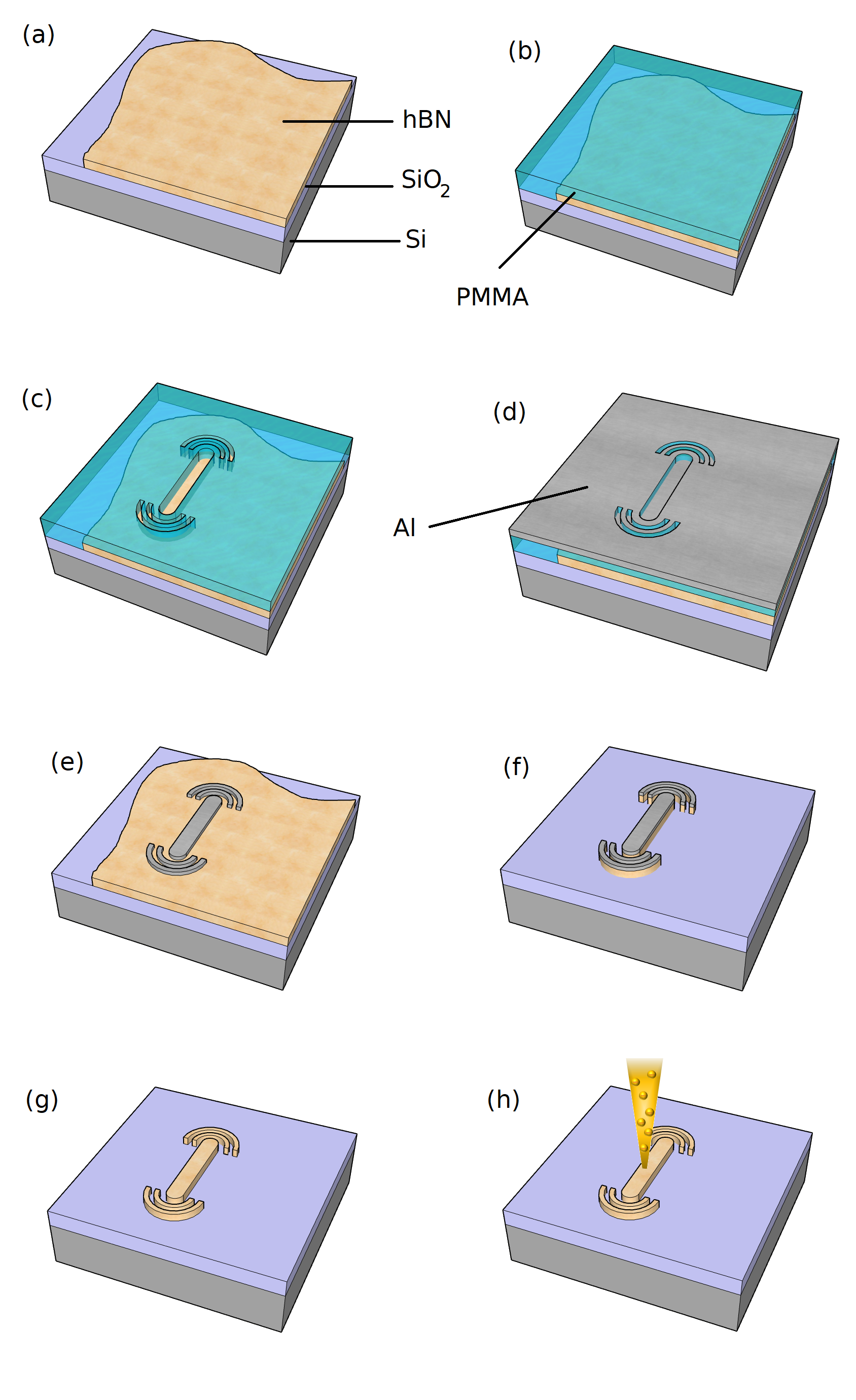}%
 \caption{\label{fig2} Fabrication of the hBN waveguide embedding quantum emitters. (a) A hBN crystal is exfoliated on a SiO$_2$/Si substrate. (b) and (c) E-beam lithography is realized on PMMA. (d) Aluminum is deposited on the sample. (e) After lift-off, the remaining Al serves as a mask. (f) The hBN flake is etched away outside of the Al mask. (g) The Al mask is removed with KOH. (h) The waveguide is irradiated to generate localized quantum emitters. }%
 \end{figure}

Figure~\ref{fig2} depicts the fabrication steps. The waveguide fabrication starts with the exfoliation of high-pressure, high-temperature grown hBN~\cite{Taniguchi07} on a SiO$_2$(300~nm)/Si substrate. Single crystals of 60 to 220~nm thickness are selected using atomic force microscopy and cathodoluminescence, to infer the quality of the crystal as well as the presence of carbon complexes, identified as precursors of the B-centers~\cite{Gale22}. The waveguides are then processed from the hBN crystals based on the following steps~\cite{Graef18}. The waveguide shape is patterned by electron beam lithography with a Raith eLine system working at 20~kV (PMMA A3, dose 250~$\mu$C/cm$^2$). We then deposit 30~nm of aluminum that, after lift-off, serves as a mask in the following step. The etching of the waveguide is performed with a fluoride reactive ion etching (RIE) for 3~min 30~s with the following parameters: plasma power of 50~W, etching pressure of 40~mTorr, 40~sccm of CHF$_3$ and 4~sccm of O$_2$ (etching speed 33~nm/minute). The aluminum is then dissolved in a KOH solution. To generate the SPEs in the fabricated waveguide, the sample is finally inserted in a SEM. The waveguide is then irradiated at precise positions located in the center of the ridge, using a static focused beam of 0.4~nA under an acceleration voltage of 15~kV during 15~s. These parameters were found to provide an average SPE yield of order one per irradiated site in this sample, based on in-situ cathodoluminescence\cite{Roux22}. The SPE generation still has a partially probabilistic character, associated with fluctuations in the SPE number, in-plane polarization direction and depth. The two latter attributes impact their coupling with the guided mode. We therefore performed four irradiations on a 60~nm thick waveguide (termed WG1) and, in the following, we focus on a SPE that presents favorable characteristics. In addition, another waveguide, denoted WG2 (thickness 220~nm), was irradiated with a higher dose to yield a localized ensemble of SPEs.

 \begin{figure}
 \includegraphics[width=0.7\linewidth]{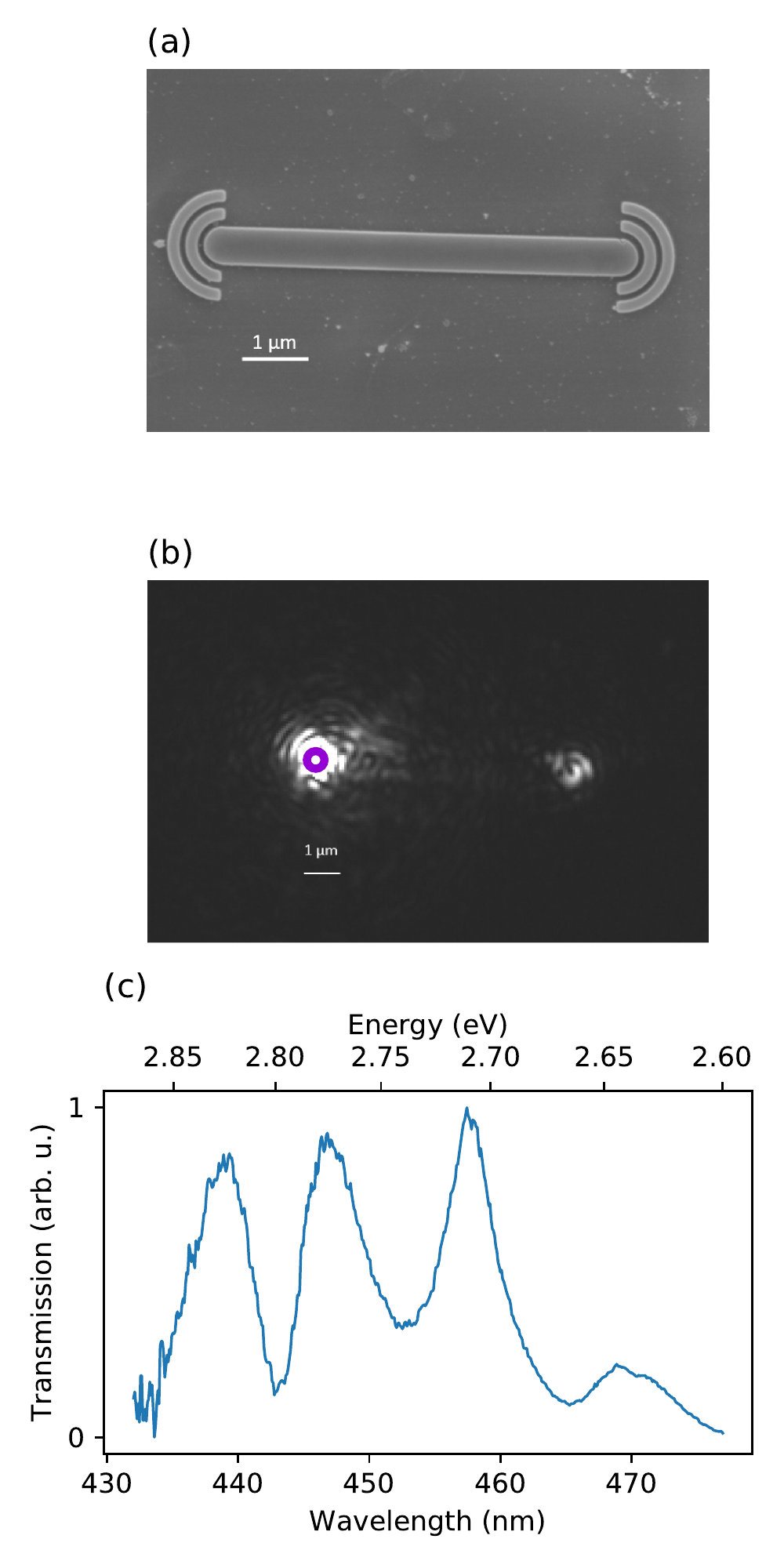}%
 \caption{\label{fig3} (a) SEM image of a waveguide. (b) CCD image of the waveguide under laser illumination focused on one of the grating couplers. The circle denotes the laser spot. (c) Transmission spectrum of a broadband source.}%
 \end{figure}
  
 \begin{figure}[t]
 \includegraphics[width=0.95\linewidth]{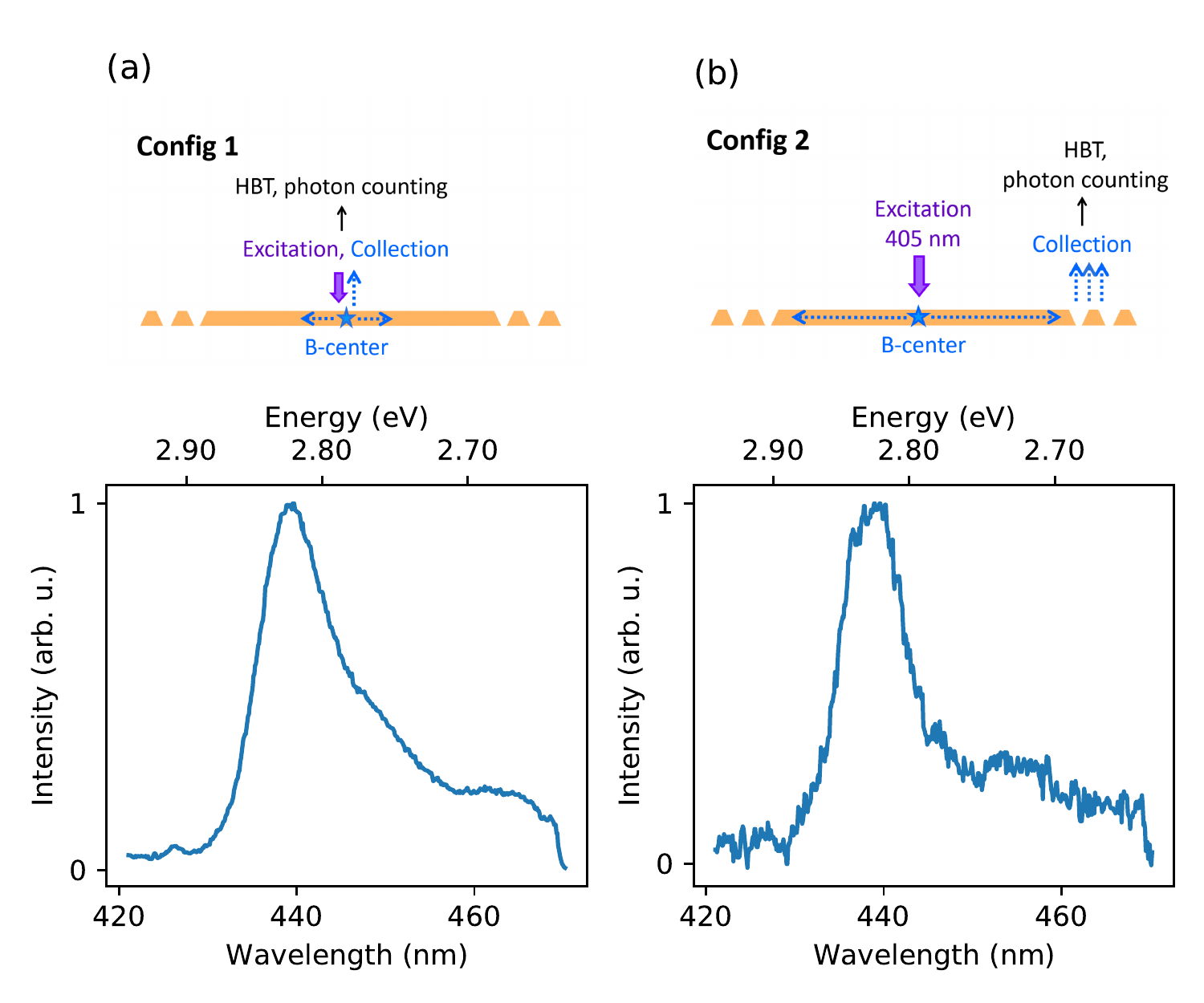}%
 \caption{\label{fig4} (a) Upper panel: Scheme of the configuration of excitation and collection path (configuration~1). Lower panel: Ensemble spectrum in configuration~1. (b) Upper panel: Scheme of configuration~2. Lower panel: Ensemble spectrum in configuration~2.}
 \end{figure}
 
 \begin{figure*}
 \includegraphics[width=0.92\linewidth]{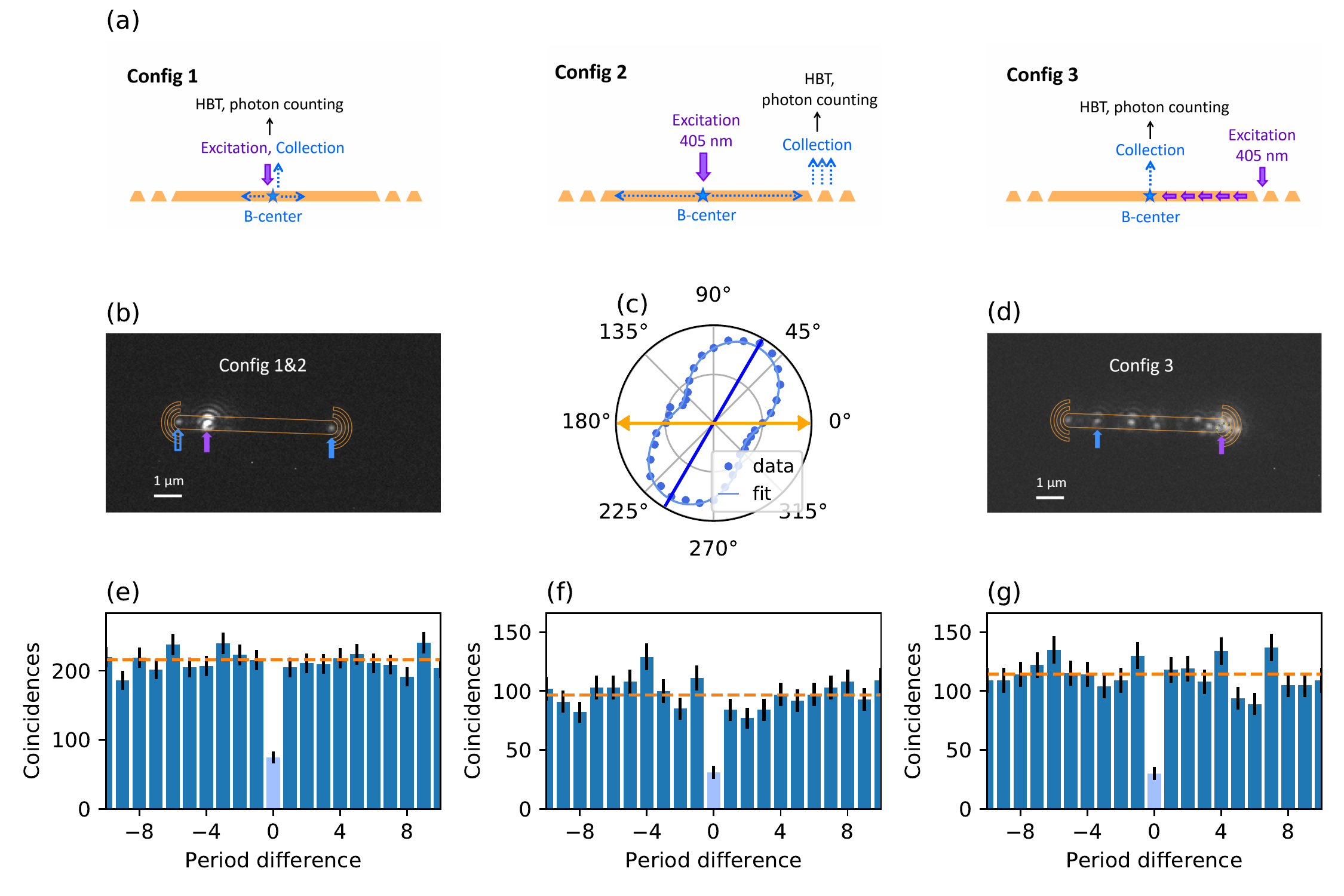}%
 \caption{\label{fig5} (a) The three configuration of excitation and collection port. (b) CCD image of the device PL in configurations 1 and 2. The violet arrow denotes the laser position (for configurations 1 and 2), as well as the collection spot in configuration~1. The plain blue arrow indicates the collection spot in configuration~2. The empty blue arrow shows photons exiting from the other collection port. (c) Polarization of the emitter with respect to the waveguide direction. The orange arrow denotes the waveguide axis and the blue line the dipole axis. (d) CCD image of the device PL in configuration 3. The violet (blue) arrows indicates the position of the laser spot (collection spot). (e), resp. (f), resp. (g): $g^{(2)}$ in configuration 1 (resp. 2, resp. 3).}
 \end{figure*}
  
A SEM image of the final structure is shown figure~\ref{fig3}(a). We characterize the waveguide in a confocal microscope operating at room temperature, equipped with a high-quantum-efficiency cooled CCD camera and avalanche photodiodes (APDs). We first verify that light can be coupled in, transmitted through and coupled out from the waveguide. Fig~\ref{fig3}(b) shows a CCD image of the waveguide under laser illumination. The presence of sizable light intensity coming from the other port demonstrates coupling from free space to the guided mode and again to free space. The waveguide transmission spectrum can be inferred from the ratio between the transmitted and the reflected spectra of a broadband laser (fig~\ref{fig3}c). It exhibits etalonning due to Fabry-Perot oscillations in the waveguide. The B-center zero-phonon line (ZPL) at 440~nm coincides with a maximum of transmission.

We then perform PL measurements. The emitters are excited with a 405~nm laser diode operating in pulsed regime (80~MHz), at a power of $\sim$400~$\mu$W, which is in the linear regime of the emitter~\cite{Fournier21}. The PL signal is filtered out from the backreflected laser using a filter centered around the emitter ZPL, and collected using either the CCD camera or the APDs. We start with WG2, where an ensemble is generated in the waveguide, to perform spectroscopy measurements. We compare two different configurations of the detection path, while exciting from the top. The configuration~1 consists in exciting and detecting via the same free-space mode, directly above the emitter (fig.~\ref{fig4}(a), upper panel). This configuration does not use the guided mode. 
In this configuration, we observe the ensemble spectrum. Its spectral shape is well known\cite{Fournier21, Roux22}, and features a 440~nm ZPL~ and phonon sidebands. We then verify that the PL light is coupled to the guided mode by switching to configuration~2, where we keep the same excitation path but we detect from one of the grating couplers, as depicted on the upper panel of figure~\ref{fig4}(b). This configuration is obtained by fixing the collection path to the chosen grating coupler, and translating the excitation beam such that it excites the emitters, as monitored by PL measured on the CCD camera. As can be seen on the lower panel of figure~\ref{fig4}(b), the spectrum is essentially unchanged by being collected through the waveguide.

In the next step, we proceed to the characterization of an individual emitter. We compare three different configurations of the excitation and detection paths, which are depicted Fig.~\ref{fig5}(a). The configurations 1 and 2 consist again in exciting directly above the emitter. Fig.~\ref{fig5}(b) shows the corresponding CCD image, with the waveguide outline superimposed for clarity. The SPE PL emission is visible at the excitation spot (violet arrow) as well as at the two output ports (blue arrows), showing that it couples to the guided mode then to free-space via the grating couplers. This coupling is enabled by the large angle between the waveguide axis and the SPE polarization axis. The latter was determined by the dependence of the count rate on the angle of a polarizer inserted in the detection port (fig~\ref{fig5}(c)). The emitter lifetime is 1.83~ns, as measured by its fluorescence decay. This value is consistent with prior measurements of B-centers in non-processed flakes~\cite{Fournier21}. Using a Hanbury Brown and Twiss setup, we measure the autocorrelation function $g^{(2)}$ of the SPE in configuration~1, where the light is directly collected from the top of the emitter, at the location depicted by the violet circle on fig.~\ref{fig5}(b). Fig~\ref{fig5}(f) shows a histogram of the photon delay times integrated over multiples of the laser repetition period. The decreased coincidence number of the center period (zero delay) with respect to the others provide $g^{(2)}(0) = 0.35 \pm 0.04$, indicating that light predominantly originates from a single B-center. This value is limited by background signal and can be largely improved by decreasing the temperature and using narrower filtering~\cite{Fournier23}. Switching to configuration 2 is done by keeping the same excitation path but detecting from one of the grating couplers (plain blue circle on fig.~\ref{fig5}(b)), as depicted on the scheme fig.~\ref{fig5}(a). In this configuration, the count rate is about a factor~4 lower, indicating that the emitter-waveguide coupling is 45~\% lower than the ideal case considered in the simulations, where the emitter is located at the mode antinode. This lower count rate could also originate from deviations of the grating coupler dimensions from the nominal values. Fig.~\ref{fig5}(e) shows the $g^{(2)}$ measured in configuration~2, which exhibits similar antibunching ($g^{(2)}(0) = 0.33 \pm 0.06$). Crucially, this demonstrates that the $g^{(2)}$ is not degraded through propagation in the structure. Finally, we show that the excitation laser can also be coupled to the guided mode (configuration~3) to excite the SPE. In this configuration, the laser excites the whole structure, such that other emitters luminesce in the waveguide and the grating couplers. Fig.~\ref{fig5}(d) shows the corresponding CCD image.  To ensure that we only detect light from the same SPE, we then collect the PL signal from the top of the waveguide, at the spot indicated by the blue arrow on fig.~\ref{fig5}(d). Fig.~\ref{fig5}(g) shows the corresponding coincidence histogram, yielding $g^{(2)}(0) = 0.26 \pm 0.04$.

Altogether, these results demonstrate that hBN fabrication and B-center generation can be combined in a complete process starting from hBN exfoliation all the way to deterministic emitter positioning. The obtained device yields guided single photons and operates at room temperature. Future improvements will require optimized photonic structures and emitter-to-photonic mode coupling and a more controlled SPE generation process.

\begin{acknowledgments}
 The authors acknowledge Christophe Arnold for his help with cathodoluminescence measurements. This work is supported by the French Agence Nationale de la Recherche (ANR) under reference ANR-21-CE47-0004-01 (E$-$SCAPE project). This work also received funding from the European Union’s Horizon 2020 research and innovation program under Grant No. 881603 (Graphene Flagship Core 3). K.W. and T.T. acknowledge support from JSPS KAKENHI (Grant Numbers 19H05790 and 20H00354).
\end{acknowledgments}

%\bibliography{your-bib-file}
~\\
%
%
%\section*{Data Availability Statement}
%The data that support the findings of this study are openly available in
%[repository name] at http://doi.org/[doi], reference number [reference number].

\end{document}